\documentclass[10pt]{llncs}

\usepackage{epsfig,amsfonts,subfigure,amsbsy,bm,mathrsfs}
\usepackage{graphicx,cite,amssymb,amsmath,color}
\usepackage{amsmath,amssymb,amscd}
\usepackage{mathtools}
\usepackage[nolist]{acronym}
\usepackage{booktabs}
\mathtoolsset{showonlyrefs=true}

\newcommand{\GL}{}
\newcommand{\HB}{}
\newcommand{\mat}[1]{\ensuremath{\bm{#1}}}
\newcommand{\F}{\ensuremath{\mathbb{F}}}
\newcommand{\weight}[1]{\mathsf{wht}\left(#1 \right)}
\renewcommand{\vec}[1]{\ensuremath{\bm{#1}}}
\newcommand{\Dec}[1]{\ensuremath{\text{DEC}_{#1}}}
\newcommand{\Imax}{\ensuremath{I_{\mathrm{max}}}}

\newcommand{\neigh}[1]{\ensuremath{\mathcal{N}\left(#1\right)}}
\newcommand{\vn}{\ensuremath{\mathsf{v}}}
\newcommand{\cn}{\ensuremath{\mathsf{c}}}
\newcommand{\msg}[2]{\ensuremath{m_{#1\rightarrow#2}}}

\newcommand{\sign}{\ensuremath{\mathrm{sign}}}
\newcommand{\degVN}{\ensuremath{d_v}}
\newcommand{\degCN}{\ensuremath{d_c}}

\newcommand{\distProfile}[1]{\ensuremath{D(#1)}}
\newcommand{\dist}{\ensuremath{d_L}}
\newcommand{\mult}[1]{\ensuremath{\mu(#1)}}
\newcommand{\errSet}[1]{\ensuremath{\Psi_{#1}}}
\newcommand{\defeq}{\ensuremath{\overset{\text{def}}{=}}}
\newcommand{\Mupc}{\ensuremath{\text{Max}_\text{upc}}}
\newcommand{\ensemble}{\ensuremath{{\mathscr{C}}}}

\newcommand{\thr}{\delta^\star}

\newcommand{\thrE}{\ensuremath{{\thr_{\mathsf{E}}}}}
\newcommand{\thrEone}{\thr_{\mathsf{1}}}
\newcommand{\thrEtwo}{\thr_{\mathsf{2}}}

\begin{document}
	
\begin{acronym}
\acro{LDPC}{low-density parity-check}
\acro{MDPC}{moderate-density parity-check}
\acro{VN}{variable node}
\acro{CN}{check node}
\acro{BP}{belief propagation}
\acro{MBP}{masked belief propagation}
\acro{BF}{bit-flipping}
\acro{EXIT}{extrinsic information transfer}
\acro{APP}{a posteriori probability}
\acro{DE}{density evolution}
\acro{RSA}{Rivest-Shamir-Adleman}
\acro{QC}{quasi-cyclic}
\acro{GJS}{Guo, Johansson and Stankovski}
\acro{MF}{Miladinovic and Fossorier}
\acro{MP}{message-passing}
\acro{REMP}{random erasure message-passing}
\acro{MI}{mutual information}
\acro{FER}{frame error rate}
\acro{BSC}{binary symmetric channel}
\acro{MAP}{maximum a posteriori}
\end{acronym}	

\title{On Decoding Schemes for the MDPC-McEliece Cryptosystem}
\author{Hannes Bartz and Gianluigi Liva
\institute{Institute of Communication and Navigation \\ Deutsches Zentrum f\"{u}r Luft- und Raumfahrt (DLR), D-82234 Wessling, Germany \\
	  {\tt \{hannes.bartz,gianluigi.liva\}@dlr.de}}}

\thispagestyle{empty} \setcounter{page}{1} \maketitle

\begin{abstract}
 \GL{Recently, it has been shown how McEliece} public-key cryptosystems based on moderate-density parity-check (MDPC) codes allow for very compact keys compared to variants based on other code families.
 In this paper, classical \GL{(iterative)} decoding schemes for MPDC codes are considered.
 The algorithms are analyzed with respect to their error-correction capability as well as their resilience against a recently proposed reaction-based key-recovery attack on a variant of the MDPC-McEliece cryptosystem by Guo, Johansson and Stankovski (GJS).
 New message-passing decoding algorithms are presented and analyzed.
 Two proposed decoding algorithms have an improved error-correction performance compared to existing hard-decision decoding schemes and are resilient against the {GJS} reaction-based attack for an appropriate choice \GL{of the algorithm's} parameters.
 \GL{Finally,} a modified belief propagation decoding algorithm that is resilient against the {GJS} reaction-based attack is presented.
\end{abstract}

\acresetall
\section{Introduction}\label{sec:intro}

In 1978,~\ac{RSA} proposed a pubilic-key cryptosystem whose security is based on the hard problem of factoring large integers.
Since then, the~\ac{RSA} cryptosystem is used in most state-of-the art communication systems and is included in many communication standards.
In 1999, Shor presented a factorization algorithm for quantum computers that is able to factor large integers in polynomial time~\cite{shor1999polynomial}. 
Thus, assuming that quantum computer of sufficient scale can be built one day, the \ac{RSA} cryptosystem can be broken in polynomial time rendering most of today's communication systems insecure.
This result gives rise to developing cryptosystems that are post-quantum secure.

In the same year as \ac{RSA}, McEliece proposed a cryptosystem based on error-correcting codes~\cite{mceliece1978public}. 
The security of the scheme relies on the hardness of decoding an unknown linear code and thus is resilient against efficient factorization attacks by quantum algorithms like Shor's algorithm.
One drawback of the scheme is the large key size and the rate-loss compared to the \ac{RSA} cryptosystem.
Variants of the McEliece cryptosystem based on different code families \GL{were considered in the past} (e.g. rank-metric codes~\cite{gabidulin1991ideals}, random codes~\cite{monico2000using}).
In particular, \GL{McEliece cryptosystems} based on \ac{LDPC} allow for very small keys but suffer from feasible attacks on the low-weight dual code due to the sparse parity-check matrix~\cite{monico2000using}.
Variants based on \ac{QC}-\ac{LDPC} codes that use row and column scrambling matrices to increase the \GL{density of the public code parity-check matrix}~\cite{baldi2016enhanced} allow for structural attacks~\cite{couvreur2015polynomial}.
The family of \ac{MDPC} codes admit a parity-check matrix of moderate density,\footnote{\GL{The existence of a moderate-density parity-check matrix for a binary linear block code does not rule out the possibility that the same code fulfills a (much) sparser parity-check matrix. As in most of the literature, we neglect the probability that a code defined by a randomly-drawn moderate parity check matrix admits a sparser parity-check matrix. Guarantees in this sense shall be derived based on random code ensemble arguments.}} \GL{yielding codes  with large minimum distance}~\cite{ouzan2009MDPC}.
In~\cite{Misoczki13:MDPC} a McEliece cryptosystem based on \ac{QC}-\ac{MDPC} codes that defeats information set decoding attacks on the dual code due to the moderate density parity-check matrix is presented.
For a given security level, the \ac{QC}-\ac{MDPC} cryptosystem allows for very small key sizes compared to other McEliece variants. 

Recently, \ac{GJS} presented a reaction-based key-recovery attack on the \ac{QC}-\ac{MDPC} system~\cite{guo2016key}.
This attack reveals the parity-check matrix by observing the decoding failure probability for chosen ciphertexts that are constructed with error patterns which have a specific structure.
A modified version of the attack can break a system that uses CCA-2 secure conversions~\cite{kobara2001semantically}. 

In this paper we analyze different decoding algorithms for (\ac{QC}-) \ac{MDPC} codes with respect to their error-correction capability and their resilience against the \ac{GJS} attack~\cite{guo2016key}. 
In particular, we present novel hard-decision \ac{MP} algorithms that are resilient against the \ac{GJS} key-recovery attack from~\cite{guo2016key} and have an improved error-correction capability compared to existing hard-decision decoding schemes.
We derive the \ac{DE} for the novel decoding schemes which allows to predict decoding thresholds as well as to optimize the parameters of the algorithm.

The paper is structured as follows. 
Section~\ref{sec:prelim} gives basic definitions, describes classical decoding schemes for \ac{LDPC}/\ac{MDPC} codes and analyzes their resilience against the \ac{GJS} attack by simulations.
In Section~\ref{sec:de} we propose new \ac{MP} decoding schemes that are able to defeat the \ac{GJS} attack. 
To estimate the decoding threshold we perform density evolution analysis of the novel schemes.
Finally, Section~\ref{sec:conclusions} concludes the paper.

\section{Preliminaries}\label{sec:prelim}

Denote the binary field by $\F_2$ and let the set of $m\times n$ matrices over $\F_2$ be denoted by $\F^{m\times n}$.
The set of all vectors of length $n$ over $\F_2$ is denoted by $\F_2^n$.
Vectors and matrices are denoted by bold lower-case and upper-case letters such as $\vec{a}$ and $\mat{A}$, respectively.
A binary circulant matrix $\bm{A}$ of size $Q$ is a $Q\times Q$ matrix with coefficients in $\F_2$ obtained by cyclically shifting its first row $\bm{a}=\left(a_0, a_1, \ldots, a_{Q-1}\right)$ to right, yielding
\begin{equation}
\bm{A}=\left(\begin{array}{cccc}
a_0 & a_1 & \cdots & a_{Q-1}\\
a_{Q-1} & a_0 & \cdots & a_{Q-2}\\
\vdots & \vdots & \ddots &\vdots\\
a_1 & a_2 & \cdots & a_{0}\\
\end{array}\right).
\end{equation} 
The set of $Q\times Q$ circulant matrices together with the matrix multiplication and addition forms a commutative ring and it is isomorphic to the polynomial ring $\left(\F_2[X]/\left(X^Q-1\right),+,\cdot\right)$. 
In particular, there is a bijective mapping between a circulant matrix $\bm{A}$ and a polynomial $a(X)=a_0+a_1X+\ldots+a_{Q-1}x^{Q-1} \in \F_2[X]$. 
We indicate the vector of coefficients of a polynomial $a(X)$ as $\bm{a}=\left(a_0, a_1, \ldots, a_{Q-1}\right)$. 
The weight of a polynomial $a(X)$ is the number of its non-zero coefficients, i.e., it is the Hamming weight of its coefficient vector $\bm{a}$. 
We indicate both weights with the operator $\weight{\cdot}$, i.e., $\weight{a(X)}=\weight{\bm{a}}$. 
In the remainder of this paper we use the polynomial representation of circulant matrices to provide an efficient description of the structure of the codes.

\subsection{QC MDPC-based Cryptosystems}

A new variant of the McEliece public-key cryptosystem that is based on \ac{QC}-\ac{MDPC} codes was proposed in~\cite{Misoczki13:MDPC}
The \ac{QC}-\ac{MDPC} McEliece cryptosystem allows for a very simple description without the need for row and column scrambling matrices.
Due to the moderate density of the parity-check matrix, \GL{known} decoding attacks on the dual code~\cite{monico2000using} are defeated.
The parity-check matrix consists of blocks of $Q\times Q$ circulant matrices which allows for very small key sizes due to the compact description of the circulant blocks.

A binary \ac{MDPC} code of length $n$, dimension $k$ and row weight $\degCN$ is defined by a binary parity-check matrix $\mat{H}$ that contains a moderate number of $\degCN\approx\mathcal{O}(\sqrt{n\log(n)})$ ones per row.
For $n=N_0Q$, dimension $k=K_0Q$, redundancy $r= n-k=R_0Q$ with $R_0=N_0-K_0$ for some integer $Q$, the parity-check matrix $\mat{H}(X)$ of a \ac{QC}-\ac{MDPC}\footnote{As in most of the recent literature on codes constructed from arrays of circulants, we loosely define a code to be \ac{QC} if there exists a permutation of its coordinates such that the resulting (equivalent) code has the following property: if $\bm{x}$ is a codeword, then any cyclic shift of $\bm{x}$ by $\ell$ positions is a codeword. For example, a code admitting a parity-check matrix as an array of \GL{$R_0\times N_0$} circulants does not fulfill the property above. However the code is \ac{QC} in the loose sense, since it is possible to permute its coordinates to obtain a code for which every cyclic shift of a codeword by \GL{$\ell=N_0$} positions yields another codeword.} code in polynomial form is a $R_0\times N_0$ matrix.

Without loss of generality we consider in the following codes with $r=Q$ (i.e. $R_0=1$). 
This family of codes covers a wide range of code rates and is of particular interest for cryptographic applications since the parity check matrices can be characterized in a very compact way.
The parity-check matrix of \ac{QC}-\ac{MDPC} codes with $r=Q$ has the form 
\begin{equation}\label{eq:H_QC-MDPC-simple}
 \mat{H}(X)=
 \begin{pmatrix}
  h_{0}(X) & h_{1}(X) & \dots & h_{N_0-1}(X)
 \end{pmatrix}.
\end{equation}
Let $\Dec{\mat{H}}(\cdot)$ be an efficient decoder for the code defined by the parity-check matrix $\mat{H}$.

\medskip
\noindent
\textbf{Key generation:}
\begin{itemize}
 \item Randomly generate a parity-check matrix $\mat{H}\in\F_2^{r\times n}$ of the form \eqref{eq:H_QC-MDPC-simple} with $\weight{h_{i}(X)}=\degCN^{(i)}$ for $i=0,\dots,N_0$.
 The matrix $\mat{H}$ with row weight $\degCN=\sum_{i=0}^{N_0-1}\degCN^{(i)}$ is the \emph{private} key.

 \item The \emph{public} key is the corresponding binary $k\times n$ generator matrix in systematic form, i.e.,
 	\begin{equation}\label{eq:genSysMatMDPC}
  	\mat{G}(X)= 
	\left(
	\begin{array}{ccc|c}
	1 & 			& & g_{0}(X) 
	 \\
	 & \ddots 	& & \vdots 
	 \\
	 & 			& 1& g_{K_0-1}(X)
	\end{array}
	\right).
\end{equation}
 	The generator matrix $\mat{G}$ can be described by $K_0Q$ bits (public key size).
\end{itemize}

\medskip
\noindent
\textbf{Encryption:}
\begin{itemize}
 \item To encrypt a plaintext $\vec{u}\in\F_2^{k}$ a user computes the ciphertext $\vec{c}\in\F_2^n$ using the public key $\mat{G}$ as
 \begin{equation}\label{eq:encryption}
  	\vec{c}=\vec{u}\mat{G}+\vec{e}
  \end{equation} 
  where $\vec{e}$ is an error vector uniformly chosen from all vectors from $\F_2^{n}$ of Hamming weight $\weight{\bm{e}}= e$.
\end{itemize}

\medskip
\noindent
\textbf{Decryption:}
\begin{itemize}
 \item To decrypt a ciphertext $\vec{c}$ the authorized recipient uses the private key $\Dec{\mat{H}}(\cdot)$ to obtain
 \begin{equation}
  	\vec{u}\mat{G}=\Dec{\mat{H}}(\vec{m}\mat{G}+\vec{e}).
 \end{equation} 
 \item Since $\mat{G}$ is in systematic form the plaintext $\vec{u}$ corresponds to the first $k$ bits of $\vec{u}\mat{G}$.
\end{itemize}

\subsection{A Reaction-Based Attack on the QC-MDPC McEliece Cryptosystem}

Beside the conventional key-recovery and decoding attacks based on information set decoding, \ac{GJS} proposed a reaction-based key-recovery attack on the \ac{QC}-\ac{MDPC} McEliece cryptosystem~\cite{Misoczki13:MDPC} which is currently the most critical attack against the scheme~\cite{sendrier2017stateOfTheArt}.
Efficient iterative decoding of \ac{LDPC}/\ac{MDPC} codes comes at the cost of decoding failures.
For \GL{example, the \ac{MDPC} codes proposed in~\cite{Misoczki13:MDPC} are operated with a target decoding failure probability lower} than $10^{-7}$.\footnote{\GL{The  \ac{MDPC} code parameters chosen in \cite{Misoczki13:MDPC} showed to empirically attain the target. An interesting question is whether a randomly generated parity-check matrix would yield the target decoding failure probability for the given set of code parameters. A possible direction to address the question is by analyzing the \ac{MDPC} code ensemble concentration properties \cite{RU01} in the finite block length regime under the given decoding algorithm.}}

The \ac{GJS} attack exploits the observation that the decoding failure probability for some particularly chosen error patterns is correlated with the structure of the secret key, i.e., the parity-check matrix $\mat{H}$.
We now describe briefly how the attack proceeds.

The \emph{Lee distance} $\dist$ between two  entries at position $i$ and $j$ of a binary vector $\vec{a}=(a_0 \ a_1 \ \dots \ a_{n-1})$ is defined as~\cite{Lee58}
 \begin{equation}
  \dist(i,j) \defeq \min\left\{|i-j|, n-|i-j|\right\}.
 \end{equation}
 The \emph{Lee distance profile}\footnote{We use the term ``Lee distance profile'' instead of ``distance spectrum'' as in~\cite{Misoczki13:MDPC} to avoid the confusion with the distance spectrum \GL{(i.e., wieght enumerator)} in Hamming metric of linear block codes.} of a binary vector $\vec{a}$ of length $Q$ is defined as
 \begin{equation}
  \distProfile{\vec{a}}\defeq \left\{d: \exists i,j \in (0,Q-1)\textrm{ s.t. } a_i=a_j=1 \textrm{ and } \dist(i,j)=d\right\}
 \end{equation}
 where the maximum distance in $\distProfile{\vec{a}}$ is $U=\lfloor\frac{Q}{2}\rfloor$.
 The \emph{multiplicity} $\mult{d}$ is defined as the number of occurrences of distance $d$ in the vector~$\vec{a}$.
 A binary vector $\vec{a}$ is fully specified by its distance profile $\distProfile{\vec{a}}$ and thus can be reconstructed \GL{with high probability} from $\distProfile{\vec{a}}$~\cite{guo2016key} (up to cyclic shifts). 

 Let $\errSet{d}$ be a set containing all binary vectors of length $n$ with exactly $t$ ones that are placed as $\lfloor\frac{t}{2}\rfloor$ pairs with Lee distance \GL{$d$} in the first $Q$ positions of the vector.
 \GL{By limiting the errors to the first $Q$ positions}, only the first circulant block $h_0(X)$ of the matrix $\mat{H}(X)$ will \GL{determine the result of} the decoding procedure. 
 The \ac{GJS} attack proceeds as follows:
\begin{itemize}
 \item \GL{For $d=1,\dots,U$} generate error sets $\errSet{d}$ of size $M$ each \GL{(with $M$ being a parameter defining, together with $U$, the number of attempts used by the attacker)}.
 \item Send $M$ ciphertexts \eqref{eq:encryption} with $\vec{e}\in\errSet{d}$ for all $d=1,\dots,U$ and \GL{measure} the \ac{FER}.
\end{itemize}
 \GL{Since the} decoding failure probability is lower for $\vec{e}\in\errSet{d}$ with $d\in\distProfile{\vec{h}_{0}}$, i.e. if \GL{$\mult{d}> 0$}, \GL{for sufficiently large $M$ the measured \ac{FER} can be used to determine the distance profile $\distProfile{\vec{h}_{0}}$}. The vector $\vec{h}_{0}$ can then be reconstructed from the distance profile $\distProfile{\vec{h}_{0}}$ using the methods from~\cite{guo2016key}.

The remaining blocks of $\mat{H}(X)$ in~\eqref{eq:H_QC-MDPC-simple}  can then be reconstructed \GL{via the generator matrix $\mat{G}(X)$} using linear algebraic relations. 
The success on the attack depends on how the systems deals with decoding failures since the \ac{FER} can only be \GL{measured if retransmissions are requested}.
Another important factor is which decoding scheme is used.
In~\cite{guo2016key,fabvsivc2017reaction} it is shown that the \ac{GJS} attack succeeds if \ac{BF} or \ac{BP} decoding algorithms are used.

In key exchange protocols the attack can be defeated by using ephemeral keys (i.e. a new key pair for every key exchange)~\cite{cake2017}.
However, this protocol-based fix can only be applied in very special scenarios.


\subsection{Classical Decoding Algorithms}\label{subsec:classicalAlgorithms}

In the following we describe classical decoding algorithms for \ac{LDPC} codes and analyze their error-correction capability for \ac{MDPC} codes as well as their resilience against the \ac{GJS} attack.
For decoding we map each ciphertext bit $c_i$ to $+1$ if $c_i=0$ and $-1$ if $c_i=1$ yielding \GL{(with some abuse of notation)} a ciphertext $\vec{c}\in\{+1,-1\}^n$.
\GL{We consider next} iterative \ac{MP} decoding on the Tanner graph \cite{Tanner81} of the code.
A Tanner graph is a bipartite graph consisting of $n$ \acp{VN} and $r$ \acp{CN}.
A \ac{VN} $\vn_j$ is connected to a \ac{CN} $\cn_i$ if the corresponding entry \GL{$h_{i,j}$} in the parity-check matrix is equal to $1$. \GL{We consider next only regular Tanner graphs, i.e, graphs for which the number of edges emanating from each \ac{VN} equals $\degVN$ and the number of edges emanating from each \ac{CN} equals $\degCN$. We refer to $\degVN$ and $\degCN$ as variable and check node degree, respectively.}
The neighborhood of a variable node \vn~is \neigh{\vn}, and similarly \neigh{\cn} denotes the neighborhood of the check node \cn.
We denote the messages from \ac{VN} $\vn_j$ to \ac{CN} $\cn_i$ by $\msg{\vn_j}{\cn_i}$ and the messages from $\cn_i$ to $\vn_j$ by $\msg{\cn_i}{\vn_j}$.
In the following we omit the indices of \acp{VN} and \acp{CN} whenever they are clear from the context. 

\medskip
\subsubsection{Bit-Flipping}\label{subsubsec:BF}

For decryption in the \ac{QC}-\ac{MDPC} cryptosystem~\cite{Misoczki13:MDPC} an efficient \ac{BF} algorithm for \ac{LDPC} codes~(see e.g.~\cite[Alg.~5.4]{RL09}) is considered. 
This algorithm is often referred to as ``Gallager's bit-flipping'' algorithm although it is \emph{different} from the algorithm proposed by Gallager in~\cite{Gallager63}.

Given a ciphertext $\vec{c}$, a threshold $b\leq r$ and a maximum number if iterations $\Imax$, the \ac{BF} algorithm proceeds as follows.
Each \ac{VN} $\vn$ is initialized with the corresponding ciphertext bit $c\in\{+1,-1\}$ and sends the message $\msg{\vn}{\cn}=c$ to all neighboring \acp{CN} $\cn\in\neigh{\vn}$.
The \acp{CN} send the messages 
\begin{equation}\label{eq:BF_CN}
  \msg{\cn}{\vn}=\prod_{\vn\GL{'} \in\neigh{\cn}}\msg{\vn\GL{'}}{\cn}
\end{equation}
to all neighboring \acp{VN} $\vn\in\neigh{\cn}$.
Note, that~\eqref{eq:BF_CN} is equivalent to the modulo two sum of all incoming messages considered over $\F_2$.
Each variable nodes counts the number of unsatisfied check equations (i.e the number of messages $\msg{\cn}{\vn}=-1$) and sends \GL{to its neighbors} the ``flipped'' ciphertext bit if at least $b$ parity-check equations are unsatisfied, i.e.
\begin{equation}\label{eq:BF_VN}
  \msg{\vn}{\cn}=
  \begin{cases}
   -c & \text{if }|\{\cn\GL{'}\in\neigh{\vn}:\msg{\cn\GL{'}}{\vn}= -1\}|\geq b 
   \\
   \phantom{-}c & \text{\GL{otherwise.}}
  \end{cases}
\end{equation}
The algorithm terminates if either all checks are satisfied or the maximum number of iterations $\Imax$ is reached.

The error-correction capability of the \ac{BF} algorithm depends on the choice of the threshold $b$.
In~\cite{huffman2010fundamentals} the threshold $b$ is selected as the maximum number of unsatisfied parity-check equations at each iteration which is denoted by $\Mupc$.
Note, that with $b=\Mupc$ the \ac{BF} algorithm is \emph{no longer} \GL{purely} a \ac{MP} algorithm on the Tanner graph of the code since $\Mupc$ has to be obtained by a global entity.

In~\cite{Misoczki13:MDPC} it is suggested to compute $b$ according to~\cite[p.~46,~Eq.~4.16]{Gallager63} which will lead to suboptimal results since the \ac{BF} decoder is different from the decoder analyzed in~\cite[Sec.~4]{Gallager63}.
To reduce the average number of iterations the threshold in~\cite{Misoczki13:MDPC} is chosen as $b=\Mupc-\delta$, where $\delta$ is a small integer that is determined empirically (see~\cite[Sec.~4]{Misoczki13:MDPC}).

\medskip
\subsubsection{Gallager B}

An efficient binary \ac{MP} decoder for~\ac{LDPC} codes, often referred to as \emph{Gallager B}, was presented an analyzed in~\cite{Gallager63}.
Each \ac{VN} $\vn$ is initialized with the corresponding ciphertext bit $c\in\{+1,-1\}$.
The \ac{VN} send the messages
\begin{equation}\label{eq:GB_VN}
 \msg{\vn}{\cn}=
 \begin{cases}
  -c & \text{if }|\{\cn'\in\neigh{\vn}\backslash\cn:\msg{\cn'}{\vn}= -c\}|\geq b 
  \\
  \phantom{-}c & \text{else}
 \end{cases}.
\end{equation}
This means that in the first iteration \ac{VN} $\vn$ sends the message $\msg{\vn}{\cn}=c$ to all neighboring \acp{CN} $\cn\in\neigh{\vn}$.
The \acp{CN} send the messages
\begin{equation}\label{eq:GB_CN}
\msg{\cn}{\vn}=\prod_{\vn'\in\neigh{\cn}\backslash\vn}\msg{\vn'}{\cn} 
\end{equation}
to the neighboring \acp{VN}.
After iterating \eqref{eq:GB_VN}, \eqref{eq:GB_CN} at most $\Imax$ times, the final decision is given by 
\begin{equation}\label{eq:GB_final}
\GL{\hat{c}}=
 \begin{cases}
  -c & \text{if }|\{\msg{\cn}{\vn}= -c\}|> b
  \\
  \phantom{-}c & \text{else}
 \end{cases}.
\end{equation}

Comparing the \ac{CN} operations \eqref{eq:BF_CN} and~\eqref{eq:GB_CN}, and the \ac{VN} operations \eqref{eq:BF_VN} and \eqref{eq:GB_VN}, one can see the before mentioned difference between the \ac{BF} algorithm an Gallager B.
\GL{For fixed $(\degVN,\degCN)$ the average error correction capability  over the \ac{BSC}  for the ensemble of  $(\degVN,\degCN)$ \ac{LDPC} codes can be analyzed, in the limit of large block lengths, using  the \ac{DE} analysis \cite{Gallager63,RU01}. Following this approach, the optimal value (in the large block length limit) for the parameter $b$ can be determined by~\cite[Eq.~4.16]{Gallager63}.}

\medskip

\subsubsection{Miladinovic-Fossorier (MF) Algorithm}\label{subsubsec:Fossorier}

Two probabilistic variants of Gallager's algorithm B that improve upon the original version were proposed by Miladinovic and Fossorier in ~\cite[Sec.~III.A]{FossorierBF2005}. \GL{We refer next to the two algorithms as \ac{MF} algorithms.}
At each iteration \GL{$\ell$} the \ac{VN} to \ac{CN} messages~\eqref{eq:GB_VN} in Gallager B are modified with a certain probability $p_{e}^{(\ell)}$.
By defining an initial value $p_e^{(0)}=p^{\ast}$ and a decrement $p_{\mathrm{dec}}\leq p^{\ast}$, one can compute $p_e^{(\ell)}$ by
\begin{equation}\label{eq:probIteration}
 p_e^{(\ell)}=
 \begin{cases}
  p_e^{(\ell -1)}-p_{\mathrm{dec}} & \text{if } p_e^{(\ell -1)}>p_{\mathrm{dec}}
  \\
  0 & \text{else}
 \end{cases}.
\end{equation}
The \acp{VN} are initialized with the corresponding ciphertext bit $c$.

\medskip
\noindent
\textbf{Variant 1 (\ac{MF}-1):}
If the number of incoming \ac{CN} messages different from $\cn$ that do not agree with $c$ exceeds the threshold $b$, i.e. if  $|\{\cn'\in\neigh{\vn}\backslash\cn:\msg{\cn'}{\vn}= -c\}|\geq b$, the \acp{VN} send the messages
\begin{equation}
\msg{\vn}{\cn}=
\begin{cases}
  -c & \text{with probability }1-p_e^{(\ell)} 
  \\
  \phantom{-}c & \text{with probability } p_e^{(\ell)} 
 \end{cases}
\end{equation}
and $\msg{\vn}{\cn}=c$ \GL{otherwise}.


\medskip
\noindent
\textbf{Variant 2 (\ac{MF}-2):}
\GL{With respect \ac{MF}-1, we shall now introduce the iteration counter for the messages that are output by \acp{VN} and by \acp{CN}. At iteration $\ell$, is the number of message at the input of a \ac{VN} $\vn$ sent by its neighboring \acp{CN}} exceeds the threshold $b$, i.e. if  $|\{\cn'\in\neigh{\vn}\backslash\cn:\msg{\cn'}{\vn}^{(\ell-1)}= -c\}|\geq b$, the \ac{VN} sends the message
\begin{equation}
\msg{\vn}{\cn}^{(\ell)} =
\begin{cases}
  -c & \text{with probability }1-p_e^{(\ell)} 
  \\
  \msg{\vn}{\cn}^{(\ell -1)} & \text{with probability } p_e^{(\ell)} 
 \end{cases}
\end{equation}
while $\msg{\vn}{\cn}^{(\ell)}=c$ \GL{otherwise}.

The check node operation as well as the final decision remains the same as in Gallager B (see~\eqref{eq:GB_CN} and \eqref{eq:GB_final}).
In general, the second variant improves upon the first variant in terms of the number of correctable errors~\cite{FossorierBF2005}.
By definition the probability $p_{e}^{(\ell)}$ has two degrees of freedom, namely $p^{\ast}$ and \GL{$p_{\mathrm{dec}}$}, which are subject to optimization.
In general there is no close form optimization of these two parameters except for using the \GL{\ac{DE}} analysis from~\cite{FossorierBF2005} as a guideline.

\medskip

\subsubsection{Algorithm E}\label{subsubsec:AlgE}

A generalization of the Gallager B algorithm that \GL{exploits} erasures, which we further refer to as \emph{Algorithm E}, was introduced and analyzed in~\cite{RU01,mitzenmacher1998note}.
To incorporate erasures the decoder requires a ternary message alphabet $\{-1,0,+1\}$, where $0$ indicates an erasure.
The \acp{VN} are initialized with the corresponding ciphertext bit $c$ and send the messages
\begin{equation}\label{eq:ALGE_VN}
\msg{\vn}{\cn}=\sign\left[\omega c+\sum_{\cn'\in\neigh{\vn}\backslash\cn}{\msg{\cn'}{\vn}}\right].
\end{equation}
\GL{Here, $\omega$ is a heuristic weighting factor that was proposed in \cite{RU01} improve the performance of Algorithm E. In \cite{RU01}  $\omega$  was allowed to change over iterations (to account for the increase of reliability of the \ac{CN} messages as the iteration number grows). We consider next the simple case where $\omega$ is kept constant through all iterations.}
The check nodes operate the same way as in Gallager B, i.e the \acp{CN} send the messages $\msg{\cn}{\vn}$ according to~\eqref{eq:GB_CN}.
After iterating \eqref{eq:GB_CN} and \eqref{eq:ALGE_VN} at most $\Imax$ times, the final decision is made as 
\begin{equation}\label{eq:ALGE_final}
\GL{\hat{c}}=\sign\left[\omega c+\sum_{\cn\in\neigh{\vn}}{\msg{\cn}{\vn}}\right].
\end{equation}
 
In~\cite{RU01} a \ac{DE} analysis for Algorithm E was derived which allows to compute an estimate of the optimal weight $\omega$.
For odd $\degVN$ Algorithm E is equivalent to Gallager B with threshold $b=\lceil\frac{\omega+\degVN-1}{2}\rceil$ and thus is also vulnerable against the \ac{GJS} attack.

\medskip

\subsubsection{Belief Propagation (BP) Decoding}\label{subsubsec:BP}
\ac{BP} decoding is \GL{a soft-decision decoding algorithm that is optimum in the \ac{MAP} sense over a cycle-free Tanner graph.}
Each \ac{VN} $\vn$ is initialized with the log-likelihood ratios
\begin{equation}
  m_{\mathrm{ch}}=c \ln \frac{n-e}{n}
\end{equation}
where $c$ is ciphertext bit corresponding to $\vn$.
The \acp{VN} send the messages 
\begin{equation}
\msg{\vn}{\cn}=m_{\mathrm{ch}}+\sum_{\cn'\in\neigh{\vn}\backslash\cn}{\msg{\cn'}{\vn}} \label{eq:SPA_VN}
\end{equation}
to the \acp{CN}.
In turn, the \acp{CN} send the messages
\begin{equation}
\msg{\cn}{\vn}=2\tanh^{-1}\left[ \prod_{\vn'\in\neigh{\cn}\backslash\vn}\tanh\left(\frac{\msg{\vn'}{\cn}}{2}\right)\right]. \label{eq:SPA_CN}
\end{equation}
After iterating \eqref{eq:SPA_VN}, \eqref{eq:SPA_CN} at most $\Imax$ times, the final decision at each \ac{VN} is made as 
\begin{equation}
\GL{\hat{c}}=\sign\left[m_{\mathrm{ch}}+\sum_{\cn\in\neigh{\vn}}{\msg{\cn}{\vn}}\right] \label{eq:SPA_final}.
\end{equation}

It was conjectured for \ac{QC}-\ac{MDPC} codes~\cite{Misoczki13:MDPC} and finally shown for \ac{QC}-\ac{LDPC} codes \cite{fabvsivc2017reaction} that the \ac{GJS} attack is also successful for \ac{QC}-\ac{MDPC} McEliece cryptosystems \GL{under} \ac{BP} decoding.

\subsection{Simulation Results}

We now present simulation results of the \ac{GJS} attack on variants of the \ac{QC}-\ac{MDPC} cryptosystem using the above described schemes.
\GL{We consider next an \ac{QC}-\ac{MDPC} code ensemble $\ensemble$ with $n=9602$ and $k=4801$ and parity-check matrix in the form 
\begin{equation}\label{eq:H_4545}
\mat{H}(X)=(h_{0}(X) \ h_{1}(X))  
\end{equation}
where $h_{0}(X)$ and $h_{1}(X)$ are two polynomials of degree less than $4801$ and $\weight{h_{0}}=\weight{h_{1}}=45$.
The ensemble $\ensemble$ was proposed in~\cite{Misoczki13:MDPC} for $80$ bit security.}
To analyze the resilience against the \ac{GJS} attack, we performed Monte Carlo simulations \GL{for codes randomly picked from $\ensemble$ collecting up to} $200$ decoding failures (frame errors) with $\Imax=50$ iterations.
For each multiplicity in $\distProfile{\vec{h}_0}$, $11$ different error sets $\errSet{d}$ (simulation points) were simulated.
As in~\cite{fabvsivc2017reaction} the weight of the error patterns was chosen such that the \ac{FER} is high enough to be \GL{easily} observable in the simulations.

\begin{figure}
\subcapcentertrue
\centering
\subfigure[]{\includegraphics[width=0.49\columnwidth]{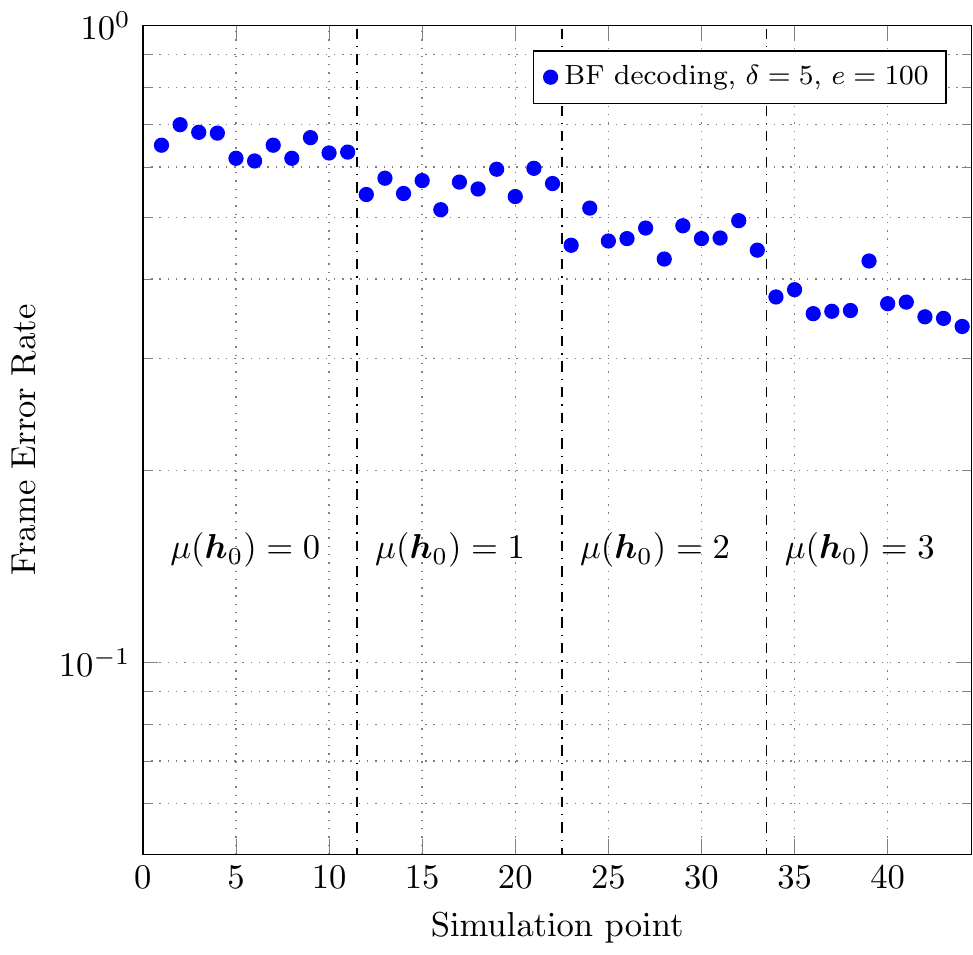}}\hfill
\subfigure[]{\includegraphics[width=0.49\columnwidth]{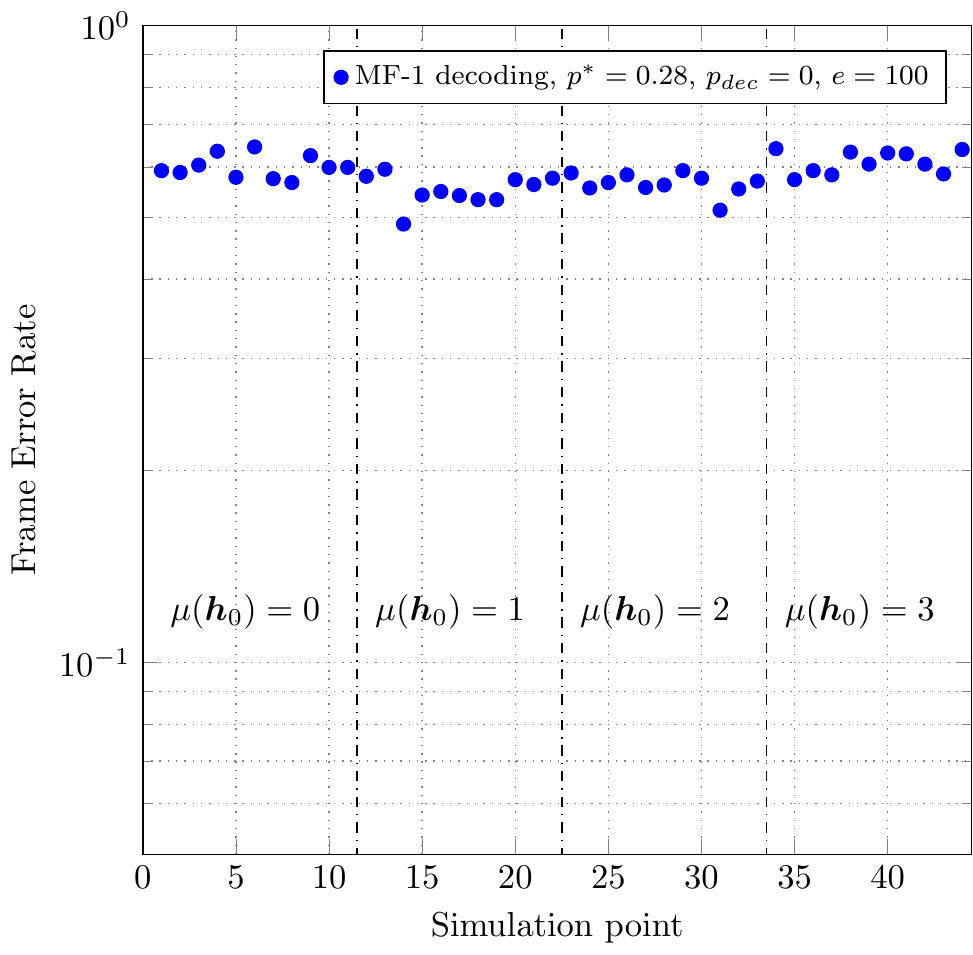}}

\subfigure[]{\includegraphics[width=0.49\columnwidth]{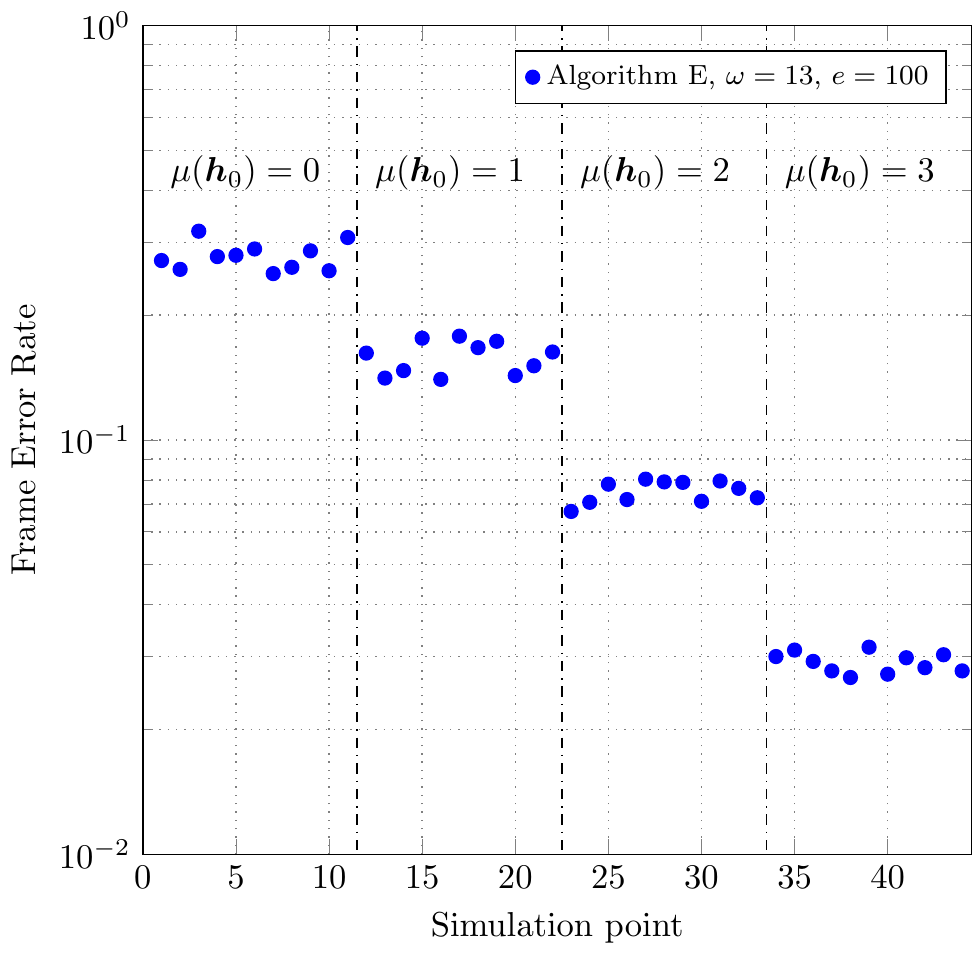}}\hfill
\subfigure[]{\includegraphics[width=0.49\columnwidth]{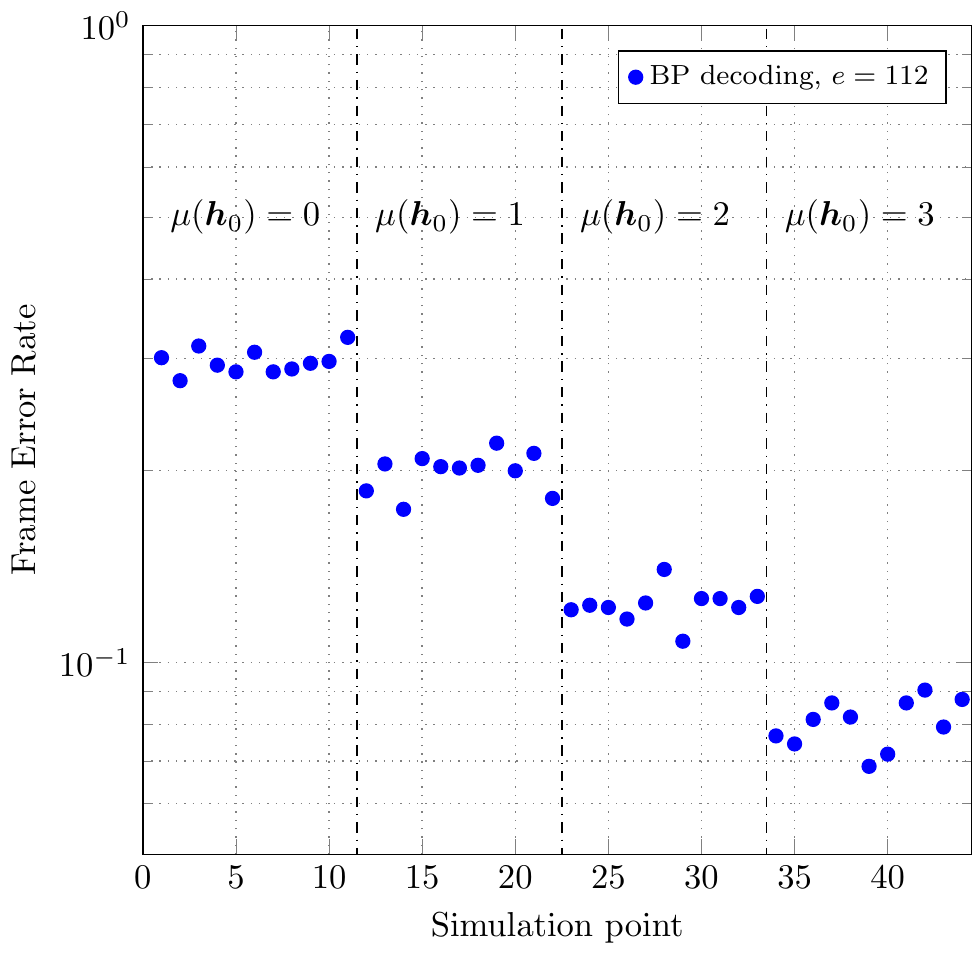}}
\caption{\ac{GJS} reaction-based attack on the code ensemble $\ensemble$ with (a) \ac{BF} decoding, (b) \ac{MF} decoding, (c) Algorithm E and (d) \ac{BP} decoding. The Monte Carlo simulation is performed with 11 simulation points per multiplicity, $I_{max}=50$ and stopping criterion of 200 decoding failures.
Except for the \ac{MF}-2 decoding scheme, the distance profile $\distProfile{\vec{h}_{0}}$ can be reconstructed from the simulation results.}
 \label{fig:reactionAttack}
\end{figure}

Figure~\ref{fig:reactionAttack} shows the simulation results for one code from $\ensemble$.
The results show that except the \ac{MF} decoding algorithm, all considered schemes are vulnerable against \ac{GJS} attack.
For the \ac{MF} decoding scheme the probability $p_e^{{(\ell)}}$ was chosen such that the \ac{FER} for all multiplicities appearing in $\distProfile{\vec{h}_0}$ are similar.
Hence, the distance profile $\distProfile{\vec{h}_{0}}$ can not be reconstructed if the \ac{MF} decoding scheme with the appropriate choice of $p_e$ is used.
Since simulations of different codes from $\ensemble$ show very similar results we \GL{conjecture} that the choice of $p_e^{(\ell)}$ rather depends on the ensemble than on the code.

\section{Secret Key Concealment via Modified Iterative Decoding}\label{sec:de}

In this section we propose new methods to modify \ac{MP} decoding algorithms that admit erasures. 
The methods allow to modify \ac{MP} decoding algorithms in a probabilistic manner to make them resilient against the \ac{GJS} attack for an appropriate choice of the decoding parameters.
The main idea is, that similar to the \ac{MF} decoding scheme (see Sec.~\ref{subsubsec:Fossorier}), we modify the \ac{VN} to \ac{CN} messages at each iteration with a given probability.
In particular, we modify the \ac{MP} decoder such that the messages $\msg{\vn}{\cn}$ are erased (\GL{i.e.,} set to $0$) under certain conditions with a given probability $p_{e}^{(\ell)}$.
\GL{Remarkably, we will see how this results also in an improved error-correction capability.}
In the following we will refer to this approach as \ac{REMP} decoding and \GL{we} apply it to modify Algorithm E.

\subsection{First Modification of Algorithm E (\ac{REMP}-1)}\label{subsec:AlgEv1}

\GL{We modify Algorithm E}  such that any nonzero message $m_{\vn\rightarrow \cn}$ in iteration $\ell$ is erased with probability $p_e^{(\ell)}$.
\GL{At the \acp{VN} we first compute a temporary output message}
\begin{equation}
\tilde{m}_{\vn\rightarrow \cn}=\sign\left[\omega c+\sum_{\cn'\in\neigh{\vn}\backslash\cn}{\msg{\cn'}{\vn}}\right].
\end{equation}
If the message $\tilde{m}_{\vn\rightarrow \cn}$ is not an erasure, i.e. if  $\tilde{m}_{\vn\rightarrow \cn}\neq 0$, the \ac{VN} sends
\begin{equation}\label{eq:ALGE3_VN}
\msg{\vn}{\cn}=
\begin{cases}
	 \tilde{m}_{\cn\rightarrow \vn} & \text{with probability }1-p_e^{(\ell)}
	 \\
	 $0$ & \text{with probability }p_e^{(\ell)}
	\end{cases}
\end{equation}
and $\msg{\vn}{\cn}=0$ else.
At the \acp{CN} we perform the same operation as in Algorithm E (see~\eqref{eq:GB_CN}).
The final decision, after iterating \eqref{eq:GB_CN} and \eqref{eq:ALGE3_VN} at most $\Imax$ times, is given by~\eqref{eq:ALGE2_final}. \GL{As for the \ac{MF} algorithm, the probability $p_e^{(\ell)}$ may be decreased as $\ell$ grows following \eqref{eq:probIteration}.}

\subsubsection{Density Evolution Analysis}

Based on the analysis of Algorithm E in~\cite{RU01}, we derive the \ac{DE} analysis of our modified algorithm from Sec.~\ref{subsec:AlgEv1}. 
Let $p_z^{(\ell)}$ denote the probability that a \ac{VN} to \ac{CN} message sent at iteration $\ell$ is equal to $z\in\{-1,0,+1\}$.
Similarly, let $q_z^{(\ell)}$ denote the probability that a \ac{CN} to \ac{VN} message sent at iteration $\ell$ is equal to $z\in\{-1,0,+1\}$.
\HB{The encryption step~\eqref{eq:encryption} can be considered as the transmission of a codeword $\vec{m}\mat{G}$ over a binary symmetric channel with crossover probability $e/n$. 
For the analysis we assume w.l.o.g. that all ciphertext bits $c_i$ are equal to $+1$ (all-zero codeword). 
Hence, we initialize the probabilities $p_{+1}^{(0)}=1-e/n$, $p_{-1}^{(0)}=e/n$ and $p_{1}^{(0)}=0$.} 

The \ac{CN} operation of \ac{REMP}-1 remains the same as in Algorithm E (see~\cite{RU01}) and thus we have
\begin{align}
 q_{+1}^{(\ell)}=&\frac{1}{2}
 \left[
 	\left(p_{+1}^{(\ell-1)}+p_{-1}^{(\ell-1)}\right)^{\degCN-1}
 	+
 	\left(p_{+1}^{(\ell-1)}-p_{-1}^{(\ell-1)}\right)^{\degCN-1}
 \right] \label{eq:q_plus}
 \\
  q_{-1}^{(\ell)}=&\frac{1}{2}
 \left[
 	\left(p_{+1}^{(\ell-1)}+p_{-1}^{(\ell-1)}\right)^{\degCN-1}
 	-
 	\left(p_{+1}^{(\ell-1)}-p_{-1}^{(\ell-1)}\right)^{\degCN-1}
 \right] \label{eq:q_minus}
 \\
 q_{0}^{(\ell)}=&1-\left(1-p_{0}^{(\ell-1)}\right)^{\degCN-1}. \label{eq:q_zero}
\end{align}
The probability $p_{+1}^{(\ell)}$ can be expressed as
\begin{align*}
 p_{+1}^{(\ell)}&=\left(1-p_e^{(\ell)}\right)\times\\ & \Bigg(
 	p_{0}^{(0)}\sum_{\substack{(i,j):\\i-j>0}}\binom{\degVN-1}{i,j,\degVN-1-i-j}
 	\left(q_{+1}^{(\ell)}\right)^{i}\left(q_{-1}^{(\ell)}\right)^{j}\left(q_{0}^{(\ell)}\right)^{\degVN-1-i-j}
 	\\
 	&+p_{+1}^{(0)}\sum_{\substack{(i,j):\\i-j>-\omega}}\binom{\degVN-1}{i,j,\degVN-1-i-j}
 	\left(q_{+1}^{(\ell)}\right)^{i}\left(q_{-1}^{(\ell)}\right)^{j}\left(q_{0}^{(\ell)}\right)^{\degVN-1-i-j}
 	\\
 	&+ p_{-1}^{(0)}\sum_{\substack{(i,j):\\i-j>\omega}}\binom{\degVN-1}{i,j,\degVN-1-i-j}
 	\left(q_{+1}^{(\ell)}\right)^{i}\left(q_{-1}^{(\ell)}\right)^{j}\left(q_{0}^{(\ell)}\right)^{\degVN-1-i-j}\Bigg).
\end{align*}
The probability $p_{-1}^{(\ell)}$ is given by
\begin{align*}
 p_{-1}^{(\ell)}&=\left(1-p_e^{(\ell)}\right)\times \\ &\Bigg(
 p_{0}^{(0)}\sum_{\substack{(i,j):\\i-j<0}}\binom{\degVN-1}{i,j,\degVN-1-i-j}
 \left(q_{+1}^{(\ell)}\right)^{i}\left(q_{-1}^{(\ell)}\right)^{j}\left(q_{0}^{(\ell)}\right)^{\degVN-1-i-j}
 \\
 &+ p_{+1}^{(0)}\sum_{\substack{(i,j):\\i-j<-\omega}}\binom{\degVN-1}{i,j,\degVN-1-i-j}
 \left(q_{+1}^{(\ell)}\right)^{i}\left(q_{-1}^{(\ell)}\right)^{j}\left(q_{0}^{(\ell)}\right)^{\degVN-1-i-j}
 \\
 &+ p_{-1}^{(0)}\sum_{\substack{(i,j):\\i-j<\omega}}\binom{\degVN-1}{i,j,\degVN-1-i-j}
 \left(q_{+1}^{(\ell)}\right)^{i}\left(q_{-1}^{(\ell)}\right)^{j}\left(q_{0}^{(\ell)}\right)^{\degVN-1-i-j}\Bigg).
\end{align*}

Finally, the probability $p_{0}^{(\ell)}$ is given by
\begin{align*}
 p_{0}^{(\ell)}=&1-p_{+1}^{(\ell)}-p_{-1}^{(\ell)}.
\end{align*}
Note that since in our scenario we do not have erasures \GL{in the ciphertext} we have $p_{0}^{(0)}=0$ which allows to simplify the expressions above.

\subsection{Second Modification of Algorithm E (\ac{REMP}-2)}\label{subsec:AlgEv2}

\GL{In the second modification of} Algorithm E from Sec.~\ref{subsubsec:AlgE} the messages $\msg{\vn}{\cn}$ at iteration $\ell$ are erased (i.e. set to $\msg{\vn}{\cn}=0$) with probability $p_e^{(\ell)}$ if they \GL{contradict} the corresponding ciphertext bit $c$.
\GL{At the \acp{VN} we first compute a temporary output message}
\begin{equation}
\tilde{m}_{\vn\rightarrow \cn}=\sign\left[\omega c+\sum_{\cn'\in\neigh{\vn}\backslash\cn}{\msg{\cn'}{\vn}}\right].
\end{equation}
If the message $\tilde{m}_{\vn\rightarrow \cn}$ \GL{contradicts} the ciphertext bit $c$, i.e. if we have \GL{$\tilde{m}_{\vn\rightarrow \cn}= -c$}, the \ac{VN} sends
\begin{equation}\label{eq:ALGE2_VN}
\msg{\vn}{\cn}=
\begin{cases}
	 \tilde{m}_{\cn\rightarrow \vn} & \text{with probability }1-p_e^{(\ell)}
	 \\
	 $0$ & \text{with probability }p_e^{(\ell)}
	\end{cases}
\end{equation}
and \GL{$\msg{\vn}{\cn}= \tilde{m}_{\cn\rightarrow \vn}$ otherwise}.
At the check nodes we perform the same operation as in Algorithm E (see~\eqref{eq:GB_CN}).
The final decision, after iterating \eqref{eq:GB_CN} and \eqref{eq:ALGE2_VN} at most $\Imax$ times, is given by 
\begin{equation}
\hat{c}=\sign\left[\omega c+\sum_{\cn\in\neigh{\vn}}{\msg{\cn}{\vn}}\right] \label{eq:ALGE2_final}.
\end{equation}
\GL{Again, as for the \ac{MF} algorithm, the probability $p_e^{(\ell)}$ may be decreased as $\ell$ grows following \eqref{eq:probIteration}.}

\subsubsection{Density Evolution}

\GL{Based on the analysis of Algorithm E in~\cite{RU01}, we derive the \ac{DE} analysis of our modified algorithm from Sec.~\ref{subsec:AlgEv2}. }

Since the \ac{CN} operation is the same as in Algorithm E, we can compute $q_{+1}^{(\ell)}, q_{-1}^{(\ell)}$ and $q_{0}^{(\ell)}$ using~\eqref{eq:q_plus}, \eqref{eq:q_minus} and~\eqref{eq:q_zero}, respectively.
The probability $p_{+1}^{(\ell)}$ can be expressed as
\begin{align*}
 p_{+1}^{(\ell)}=&
 	p_{0}^{(0)}\sum_{\substack{(i,j):\\i-j>0}}\binom{\degVN-1}{i,j,\degVN-1-i-j}
 	\left(q_{+1}^{(\ell)}\right)^{i}\left(q_{-1}^{(\ell)}\right)^{j}\left(q_{0}^{(\ell)}\right)^{\degVN-1-i-j}
 	\\
 	+&p_{+1}^{(0)}\sum_{\substack{(i,j):\\i-j>-\omega}}\binom{\degVN-1}{i,j,\degVN-1-i-j}
 	\left(q_{+1}^{(\ell)}\right)^{i}\left(q_{-1}^{(\ell)}\right)^{j}\left(q_{0}^{(\ell)}\right)^{\degVN-1-i-j}
 	\\
 	+& \left(1-p_e^{(\ell)}\right) p_{-1}^{(0)}\sum_{\substack{(i,j):\\i-j>\omega}}\binom{\degVN-1}{i,j,\degVN-1-i-j}
 	\left(q_{+1}^{(\ell)}\right)^{i}\left(q_{-1}^{(\ell)}\right)^{j}\left(q_{0}^{(\ell)}\right)^{\degVN-1-i-j}.
\end{align*}
The probability $p_{-1}^{(\ell)}$ is given by
\begin{align*}
 p_{-1}^{(\ell)}=&1-p_{+1}^{(\ell)}-p_{0}^{(\ell)}
 \\
 =&
 p_{0}^{(0)}\sum_{\substack{(i,j):\\i-j<0}}\binom{\degVN-1}{i,j,\degVN-1-i-j}
 \left(q_{+1}^{(\ell)}\right)^{i}\left(q_{-1}^{(\ell)}\right)^{j}\left(q_{0}^{(\ell)}\right)^{\degVN-1-i-j}
 \\
 +& \left(1-p_e^{(\ell)}\right) p_{+1}^{(0)}\sum_{\substack{(i,j):\\i-j<-\omega}}\binom{\degVN-1}{i,j,\degVN-1-i-j}
 \left(q_{+1}^{(\ell)}\right)^{i}\left(q_{-1}^{(\ell)}\right)^{j}\left(q_{0}^{(\ell)}\right)^{\degVN-1-i-j}
 \\
 +& p_{-1}^{(0)}\sum_{\substack{(i,j):\\i-j<\omega}}\binom{\degVN-1}{i,j,\degVN-1-i-j}
 \left(q_{+1}^{(\ell)}\right)^{i}\left(q_{-1}^{(\ell)}\right)^{j}\left(q_{0}^{(\ell)}\right)^{\degVN-1-i-j}.
\end{align*}
Finally, the probability $p_{0}^{(\ell)}$ can then be expressed as
\begin{align*}
 p_{0}^{(\ell)}=&1-p_{+1}^{(\ell)}-p_{-1}^{(\ell)}.
\end{align*}
\GL{As before, note that since in our scenario we do not have erasures {in the ciphertext} we have $p_{0}^{(0)}=0$ which allows to simplify the expressions above.}

\subsection{Masked Belief Propagation (MBP) Decoding}

Using the ideas from the \ac{MF} algorithm we now modify the classical \ac{BP} decoding algorithm (see~Sec.~\ref{subsubsec:BP}) in order to \GL{counteact} the \ac{GJS} attack.
We set
\begin{equation}
  m_{\mathrm{ch}}=c \ln \frac{n-e}{n}
\end{equation}
where $c$ is ciphertext bit corresponding to $\vn$.
The \acp{VN} first compute the \GL{temporary} messages 
\begin{equation}
\tilde{m}_{\vn\rightarrow \cn}=m_{\mathrm{ch}}+\sum_{\cn'\in\neigh{\vn}\backslash\cn}{\msg{\cn'}{\vn}}.
\end{equation}
If the sign of $\tilde{m}_{\vn\rightarrow \cn}$ is not equal to the sign of $m_{\mathrm{ch}}$, i.e. if $\sign(\tilde{m}_{\vn\rightarrow \cn})\neq\sign(m_{\mathrm{ch}})$, then the \ac{VN} sends the message
\begin{equation}\label{eq:SPA_VN_2}
\msg{\vn}{\cn}=
\begin{cases}
	 \tilde{m}_{\vn\rightarrow \cn} & \text{with probability }1-p_e^{(\ell)}
	 \\
	 m_{\mathrm{ch}} & \text{with probability }p_e^{(\ell)}
	\end{cases}
\end{equation}
and $\msg{\vn}{\cn}=\tilde{m}_{\vn\rightarrow \cn}$ \GL{otherwise}.
In other words, if the sign of a message that is supposed so be sent by \ac{VN} $\vn$ is different from the sign of the corresponding initial value $m_{\mathrm{ch}}$, then with probability $p_e^{(\ell)}$ the initial value $m_{\mathrm{ch}}$ is sent.
The \acp{CN} operation remains the same as in~\eqref{eq:SPA_CN}.
After iterating \eqref{eq:SPA_VN_2}, \eqref{eq:SPA_CN} at most $\Imax$ times, the final decision at each \ac{VN} is made according to~\eqref{eq:SPA_final}. \GL{For \ac{MBP} decoding we do not provide an explicit description on how \ac{DE} has to be modified since the analysis can be carried out by applying minor changes to quantized \ac{DE} \cite{chung2001design}.}

\GL{We shall see next that,} due to the modified operation at the \acp{VN}, the \ac{MBP} algorithm allows to conceal the structure of $\mat{H}(X)$ by tuning the probability $p_e^{(\ell)}$.
\GL{We empirically verified that} the idea of introducing random erasures as in~Sec.~\ref{subsec:AlgEv1} and Sec.~\ref{subsec:AlgEv2} does not conceal the structure of $\mat{H}(X)$ for \ac{BP} decoding.
\GL{Moreover, we will see that, differently from} the \ac{REMP} modifications of Algorithm E, the modification of \ac{BP} decoding comes at the cost of a reduced error correction performance.
Thus, the decoding algorithms from Sec.~\ref{subsec:AlgEv1} and Sec.~\ref{subsec:AlgEv2}  are preferable since they show a similar performance at a lower decoding complexity.

\subsection{Performance Analysis \& Simulation Results}

\subsubsection{Density Evolution Analysis}

We \GL{first analyze} the error-correction capability of the two modifications of Algorithm E from~Sec~\ref{subsec:AlgEv1} and Sec.~\ref{subsec:AlgEv2}. \GL{As first estimate of the code performance, we employ the \ac{DE} analysis \cite{RU01} to determine the iterative decoding threshold of a $(\degVN,\degCN)$ unstructured \ac{LDPC} code ensemble over a \ac{BSC} with error probability $\Delta$. The decoding threshold is denoted as $\Delta^\star$ and represents the largest channel error probability for which, in the limit of large $n$ and large $\Imax$, the bit error probability of code picked randomly from the ensemble becomes vanishing small \cite{RU01}. We then get a rough estimate on the error correction capability as\footnote{\GL{Note that at the decoding threshold $\Delta^\star$ a vanishing small bit error probability may not imply a vanishing small block error probability. However, for the regular \ac{MDPC} ensembles under consideration the threshold on the bit error probability and the one on the block error probability do coincide over  binary-input output-symmetric memoryless channel under \ac{BP} decoding \cite{Lentmaier05}. In our estimate, we implicitly assume that the result extends to Algorithm E and its variants.}}} 
\[	
	\delta^\star=\lfloor n\Delta^\star\rfloor.
\]
\GL{Note that, for a moderate block length $n$, $\delta^\star$ provides only a coarse estimate to the number of errors at which we expect the \ac{FER} to rapidly decrease (so-called waterfall region), with the accuracy of the prediction improving as $n$ grows large. With a slight abuse of the wording, we refer to $\delta^\star$ as decoding threshold as well.}
We \GL{further} denote the decoding threshold under Algorithm E, \ac{REMP}-1 and \ac{REMP}-2 as $\thrE$, $\thrEone$ and $\thrEtwo$,  respectively. \GL{The decoding thresholds do not only depend on the selected algorithm, but also on the algorithm parameters. The results for the $(9602,4801)$ \ac{MDPC} ensemble with $\degVN=45$ and $\degCN=90$ are summarized in Table \ref{tab:decThresholdsSecLevels}. For Algorithm E, the value of $\omega$ has been chosen to maximize the decoding threshold. Remarkably, the variants \ac{REMP}-1 and \ac{REMP}-2 do not yield a threshold degradation, and in some cases they even provide slight gains for suitable choices of the parameters $(\omega,p^{\ast},p_{\mathrm{dec}})$.}

\begin{table}[ht!]
	\caption{Decoding thresholds of Algorithm E and it variants for the \ac{MDPC} code ensembles with the parameters from~\cite[Tab.~2]{Misoczki13:MDPC}.}
	\label{tab:decThresholdsSecLevels}
\centering
\begin{tabular}{|c||c|c|c||c|c|c||c|c|c||c|}
 \cline{5-11}
   \multicolumn{4}{}{}  & \multicolumn{3}{|c||}{REMP-1} & \multicolumn{3}{|c||}{REMP-2} & Alg. E  \\\hline
 Security Level & $n$ & \degCN & \degVN & $p^{\ast}$ & $p_{\mathrm{dec}}$ & $\thrEone$ ($\omega$) & $p^{\ast}$ & $p_{\mathrm{dec}}$ &  $\thrEtwo$ ($\omega$) & $\thrE$ ($\omega$)  \\\hline\hline
 80 & 9602 & 90 & 45 & 0.001 & 0 & 107(13) & 0.1 & 0 & 108(13) & 106(14)  
 \\
 128 & 19714 & 142 &  71 & 0.1 & 0.001  & 153(18) & 0.76 & 0 & 157(14) & 153(18)  
 \\
 256 & 65542 & 274  &  137 & 0.002  & 0.0002 & 296(27) & 0.65 & 0 & 301(23) & 294(26)  
 \\\hline
\end{tabular}
\end{table}

\subsubsection{Simulation Results}

\begin{figure}[ht!]
\centering
 \includegraphics[width=0.9\columnwidth]{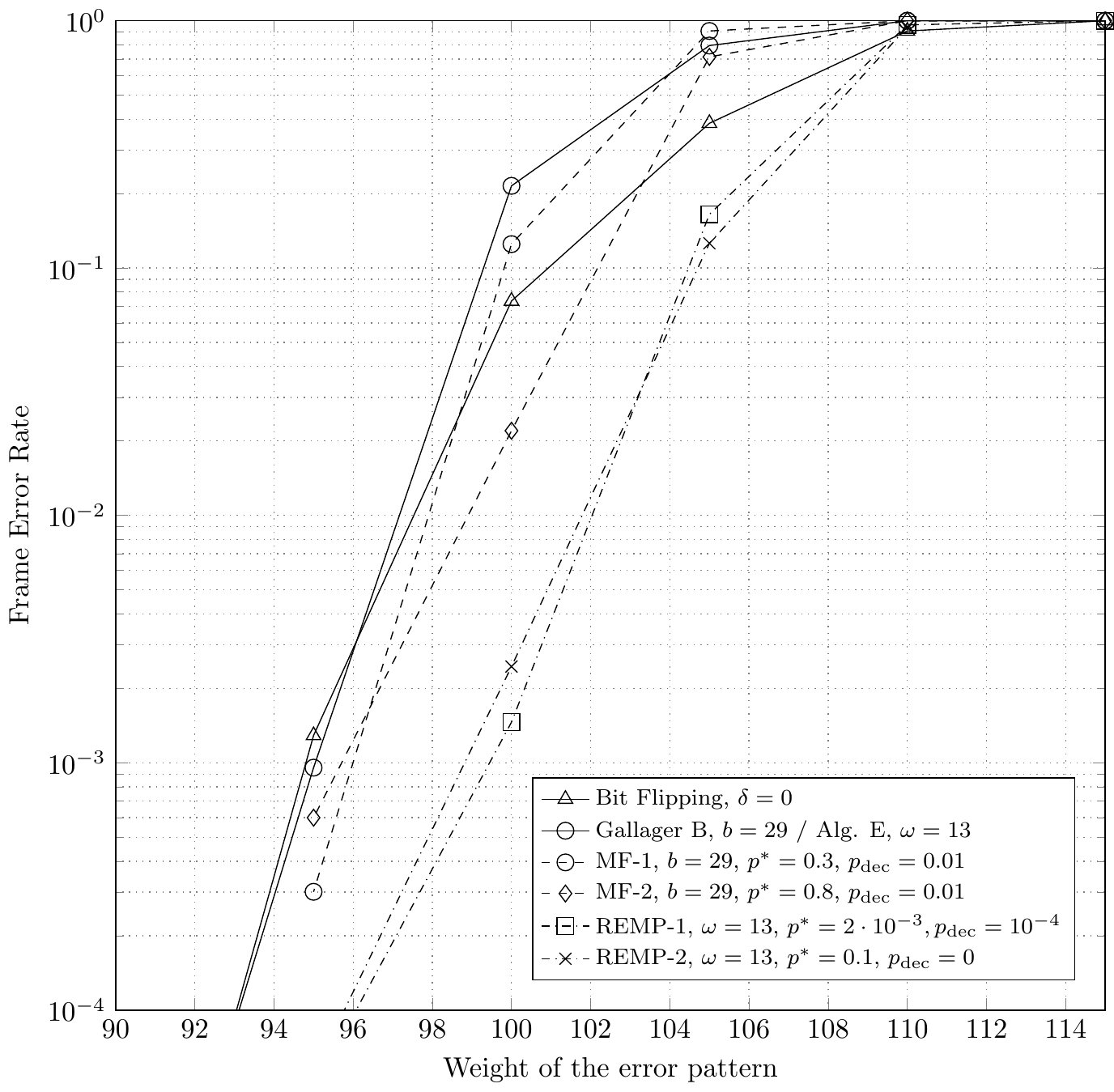}
 \caption{Error-correction performance (\ac{FER}) over the weight of the error patterns. The figure shows that the proposed \ac{REMP} schemes significantly improve upon existing hard-decision decoding schemes.}
 \label{fig:errorCurves}
\end{figure}

To validate the performance estimates \GL{obtained through \ac{DE}}, we simulated the error-correction capability of the decoding schemes from Section~\ref{subsec:classicalAlgorithms} and Section~\ref{sec:de}. The results \GL{in terms of \ac{FER} as a function of the error pattern weight are depicted} in Figure~\ref{fig:errorCurves}. \GL{The results confirm the trend predicted by the \ac{DE} analysis. In particular, the error-correction capability improves upon existing decoding algorithms. Even for erasure probability values chosen to conceal the structure of $\mat{H}(X)$ (yielding a suboptimal choice with respect to the error correction performance),  \ac{REMP}-2 outperforms Algorithm E and the \ac{BF}/\ac{MF} algorithms.}

\subsection{Resilience against the GJS Attack}

\begin{figure}
\subcapcentertrue
\centering
\subfigure[]{\includegraphics[width=0.49\columnwidth]{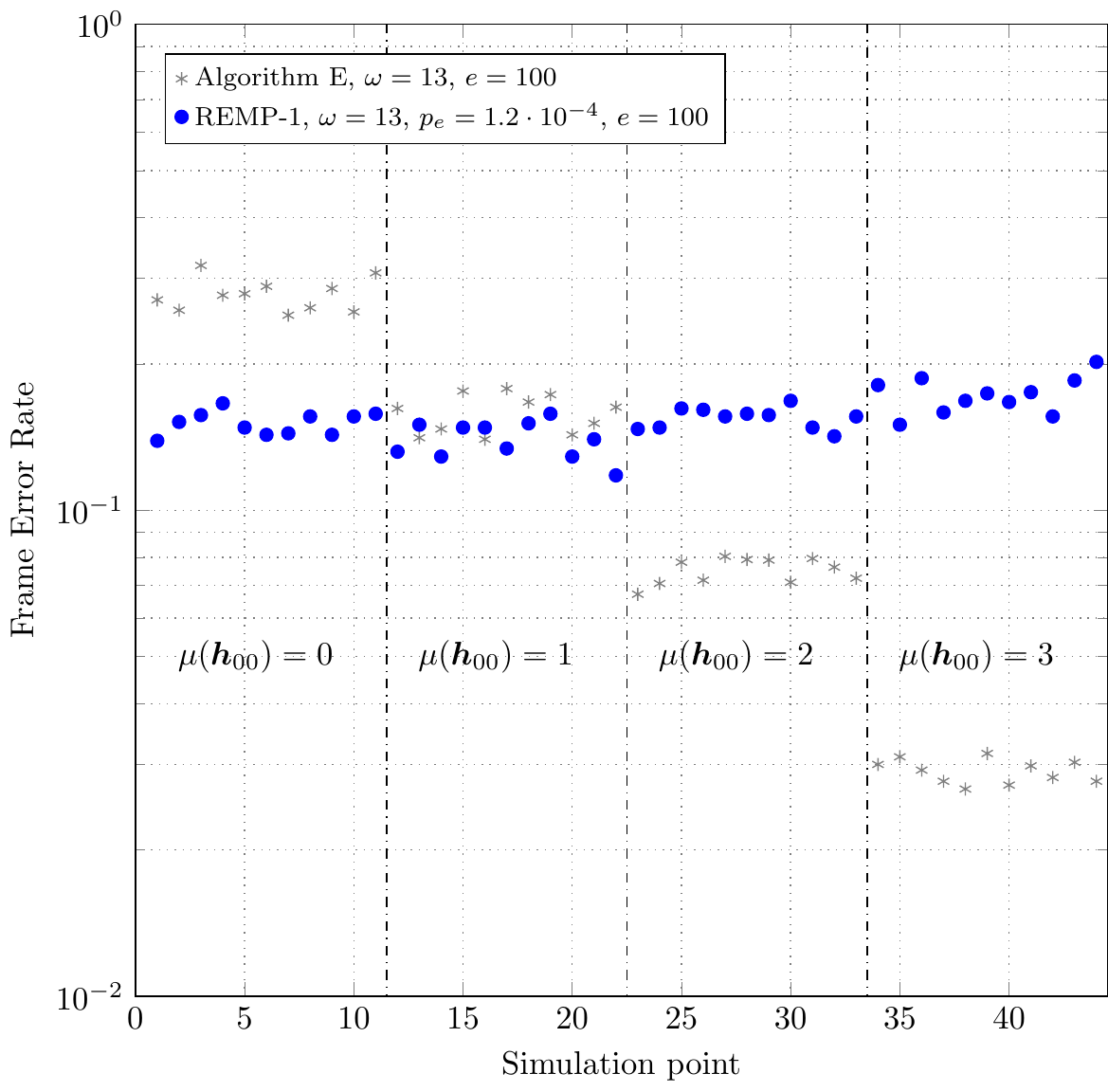}}\hfill
\subfigure[]{\includegraphics[width=0.49\columnwidth]{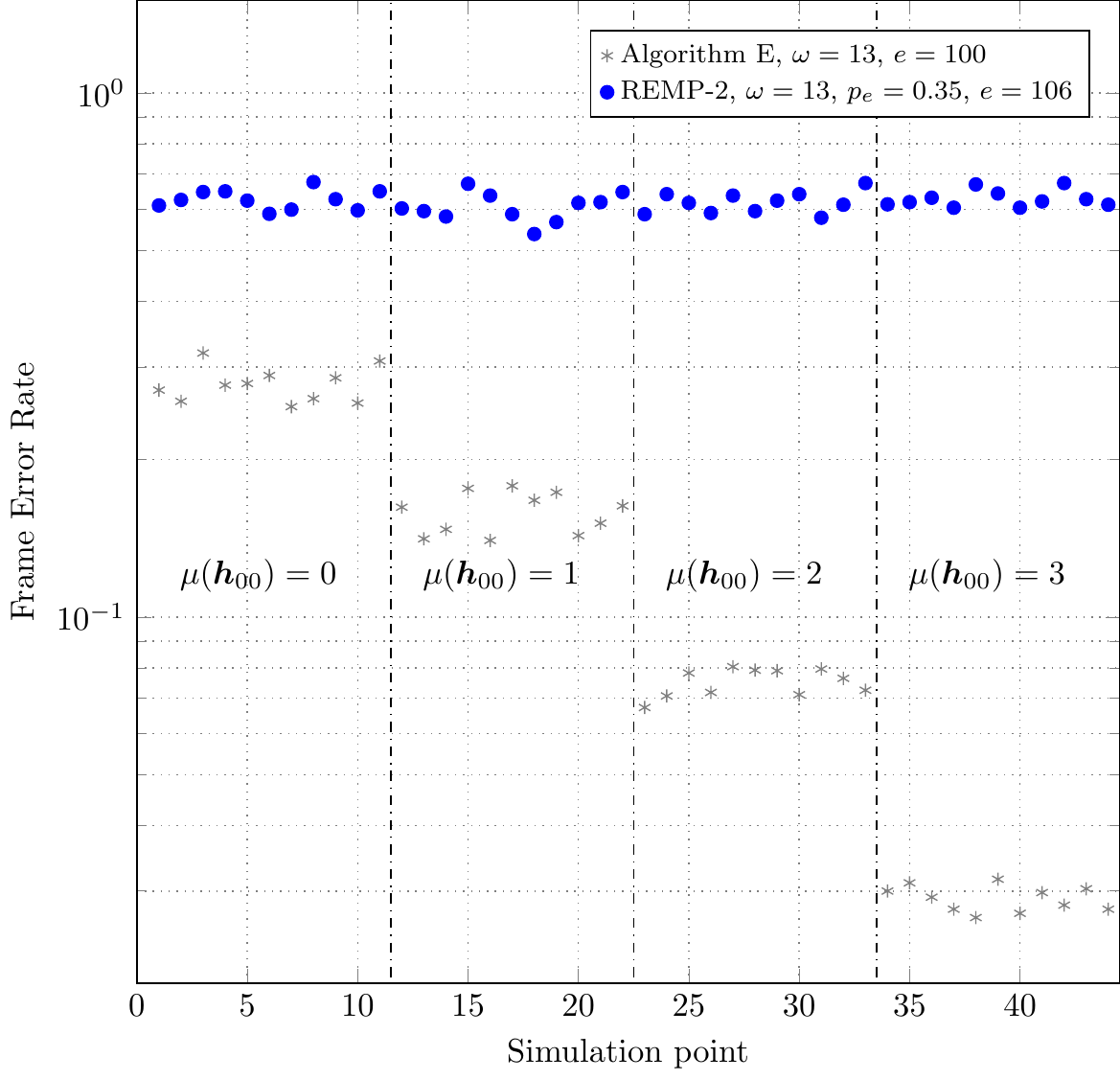}}

\caption{\ac{GJS} reaction-based attack on the code ensemble $\ensemble$ with (a) decoding, (b) decoding. The Monte Carlo simulation is performed with 11 simulation points per multiplicity, $I_{max}=50$ and stopping criterion of 200 decoding failures.
The results show that the distance profile $\distProfile{\vec{h}_{0}}$ cannot be reconstructed from the simulation results.}
 \label{fig:reactionAttackModified}
\end{figure}

We now analyze the resilience of the proposed decoding schemes against the \ac{GJS} attack. 
For the \ac{REMP} variants of Algorithm E as well as for the \ac{MBP} decoder we performed Monte Carlo simulations for codes randomly picked from $\ensemble$ collecting up to $200$ decoding failures (frame errors) with $\Imax=50$ iterations.
For each multiplicity in $\distProfile{\vec{h}_0}$, $11$ different error sets $\errSet{d}$ (simulation points) were simulated.
The simulation results in~Figure~\ref{fig:reactionAttackModified} show that, for an appropriate choice of parameters, the \ac{REMP}-1 and \ac{REMP}-2 decoding schemes have a similar~\ac{FER} for all multiplicities appearing in $\distProfile{\vec{h}_0}$.
Hence, the reconstruction of the distance profile $\distProfile{\vec{h}_0}$ from the observed \ac{FER} is not possible. 

\begin{figure}[ht!]
\centering
 \includegraphics[width=0.7\columnwidth]{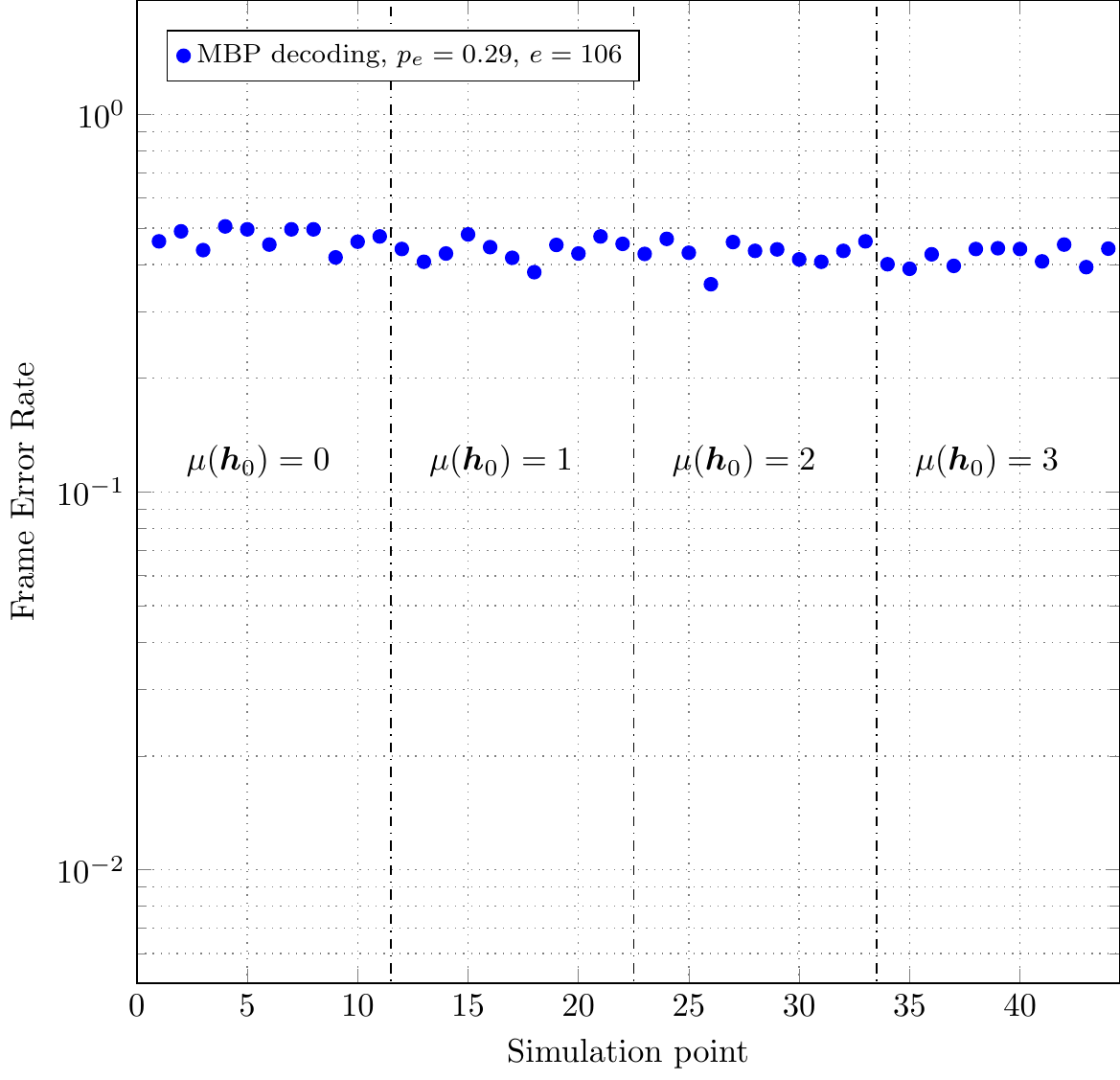}
 \caption{\ac{GJS} reaction-based attack on the code ensemble $\ensemble$ with \ac{MBP} decoding.}
 \label{fig:reactionAttackPBP}
\end{figure}

Figure~\ref{fig:reactionAttackPBP} shows that for an appropriate choice of parameters also the \ac{MBP} algorithm is able to conceal the structure of the secret key. 
For the choice of parameters that conceal the secret key the \ac{FER} of \ac{MBP} decoding and \ac{REMP}-2 decoding at error weight $e=106$ is similar.
Hence, due to the higher complexity of \ac{MBP}, the \ac{REMP} scheme is preferable.

To conceal the structure of $\mat{H}(X)$ the choice of $p_e^{(\ell)}$ for a particular error weight $e$ is crucial. 
If $p_e^{(\ell)}$ is chosen too large the picture is inverted, i.e. higher multiplicities have a higher \ac{FER} than lower multiplicities.
Thus the error weight $e$ should be computed after decoding and ciphers generated with an error weight different from $e$ should be rejected to prevent attacks that exploit this effect.

\section{Conclusions}\label{sec:conclusions}
\acresetall

We analyzed classical iterative decoding schemes for \ac{MDPC} codes with respect to their error-correction capability as well as their resilience against a recent key-recovery attack by \ac{GJS}.
The simulation results show that a decoding scheme by \ac{MF} is able to defeat the attack for an appropriate choice of decoding parameters.

A new decoding method called \ac{REMP} that allows to improve existing \ac{MP} decoding algorithms with respect to their error-correction capability as well as their resilience against the \ac{GJS} attack was proposed.
Two \ac{REMP} variants of an existing \ac{MP} decoder that have an improved error-correction performance for \ac{MDPC} codes compared to existing schemes were presented and analyzed.
The simulation results show that the proposed \ac{REMP} schemes are able to defeat the \ac{GJS} attack for an appropriate choice of decoding parameters.

A new variant of the belief propagation decoding algorithm that is able to resist the \ac{GJS} attack was presented.


\begin{thebibliography}{10}
\providecommand{\url}[1]{#1}
\csname url@samestyle\endcsname
\providecommand{\newblock}{\relax}
\providecommand{\bibinfo}[2]{#2}
\providecommand{\BIBentrySTDinterwordspacing}{\spaceskip=0pt\relax}
\providecommand{\BIBentryALTinterwordstretchfactor}{4}
\providecommand{\BIBentryALTinterwordspacing}{\spaceskip=\fontdimen2\font plus
\BIBentryALTinterwordstretchfactor\fontdimen3\font minus
  \fontdimen4\font\relax}
\providecommand{\BIBforeignlanguage}[2]{{%
\expandafter\ifx\csname l@#1\endcsname\relax
\typeout{** WARNING: IEEEtran.bst: No hyphenation pattern has been}%
\typeout{** loaded for the language `#1'. Using the pattern for}%
\typeout{** the default language instead.}%
\else
\language=\csname l@#1\endcsname
\fi
#2}}
\providecommand{\BIBdecl}{\relax}
\BIBdecl

\bibitem{shor1999polynomial}
P.~W. Shor, ``{Polynomial-Time Algorithms for Prime Factorization and Discrete
  Logarithms on a Quantum Computer},'' \emph{SIAM J. Comput.}, vol.~26, no.~5,
  pp. 1484--1509, 1997.

\bibitem{mceliece1978public}
R.~J. McEliece, ``{A Public-Key Cryptosystem Based on Algebraic Codes},''
  \emph{Deep Space Network Progress Report}, vol.~44, pp. 114--116, 1978.

\bibitem{gabidulin1991ideals}
E.~M. Gabidulin, A.~Paramonov, and O.~Tretjakov, ``{Ideals Over a
  Non-Commutative Ring and Their Application in Cryptology},'' in \emph{10th
  Annual International Conference on Theory and Application of Cryptographic
  Techniques (EUROCRYPT)}, Brighton, UK, Apr. 1991, pp. 482--489.

\bibitem{monico2000using}
C.~Monico, J.~Rosenthal, and A.~Shokrollahi, ``{Using Low Density Parity Check
  Codes in the McEliece Cryptosystem},'' in \emph{Proc. {IEEE} Int. Symp. Inf.
  Theory (ISIT)}, Sorrento, Italy, Jun. 2000, p. 215.

\bibitem{baldi2016enhanced}
M.~Baldi, M.~Bianchi, F.~Chiaraluce, J.~Rosenthal, and D.~Schipani, ``{Enhanced
  Public Key Security for the McEliece Cryptosystem},'' \emph{Journal of
  Cryptology}, vol.~29, no.~1, pp. 1--27, Jan. 2016.

\bibitem{couvreur2015polynomial}
A.~Couvreur, A.~Otmani, J.-P. Tillich, and V.~Gauthier-Umana, ``{A
  Polynomial-Time Attack on the BBCRS Scheme},'' in \emph{Proc. 18th IACR
  International Conference on Practice and Theory in Public-Key Cryptography},
  Gaithersburg, MD, USA, Mar. 2015, pp. 175--193.

\bibitem{ouzan2009MDPC}
S.~Ouzan and Y.~Be'ery, ``{Moderate-Density Parity-Check Codes},'' \emph{arXiv
  preprint arXiv:0911.3262}, 2009.

\bibitem{Misoczki13:MDPC}
R.~Misoczki, J.~P. Tillich, N.~Sendrier, and P.~S. L.~M. Barreto,
  ``{MDPC-McEliece: New McEliece variants from Moderate Density Parity-Check
  codes},'' in \emph{Proc. {IEEE} Int. Symp. Inf. Theory (ISIT)}, Istanbul,
  Turkey, Jul. 2013, pp. 2069--2073.

\bibitem{guo2016key}
Q.~Guo, T.~Johansson, and P.~Stankovski, ``{A Key Recovery Attack on MDPC with
  CCA Security Using Decoding Errors},'' in \emph{22nd Annual International
  Conference on the Theory and Applications of Cryptology and Information
  Security (ASIACRYPT)}, Hanoi, Vietnam, Dec. 2016, pp. 789--815.

\bibitem{kobara2001semantically}
K.~Kobara and H.~Imai, ``{Semantically secure McEliece public-key
  cryptosystems-conversions for McEliece PKC},'' in \emph{4th International
  Workshop on Practice and Theory in Public Key Cryptography (PKC)}, Cheju
  Island, South Korea, Feb. 2001, pp. 19--35.

\bibitem{sendrier2017stateOfTheArt}
N.~Sendrier, ``{Code-Based Cryptography: State of the Art and Perspectives},''
  \emph{IEEE Security \& Privacy}, vol.~15, no.~4, pp. 44--50, Aug. 2017.

\bibitem{RU01}
T.~Richardson and R.~Urbanke, ``{The Capacity of Low-Density Parity-Check Codes
  Under Message-Passing Decoding},'' \emph{{IEEE} Trans. Inf. Theory}, vol.~47,
  no.~2, pp. 599 -- 618, Feb. 2001.

\bibitem{Lee58}
C.~Lee, ``{Some Properties of Nonbinary Error-Correcting Codes},'' \emph{IRE
  Transactions on Information Theory}, vol.~4, no.~2, pp. 77--82, Jun. 1958.

\bibitem{fabvsivc2017reaction}
T.~Fab{\v{s}}i{\v{c}}, V.~Hromada, P.~Stankovski, P.~Zajac, Q.~Guo, and
  T.~Johansson, ``{A Reaction Attack on the QC-LDPC McEliece Cryptosystem},''
  in \emph{International Workshop on Post-Quantum Cryptography}, 2017, pp.
  51--68.

\bibitem{cake2017}
P.~S. L.~M. Barreto, S.~Gueron, T.~G\"ueneysu, R.~Misoczki, E.~Persichetti,
  N.~Sendrier, and J.-P. Tillich, ``{CAKE: Code-Based Algorithm for Key
  Encapsulation},'' Cryptology ePrint Archive, Report 2017/757, 2017,
  \url{http://eprint.iacr.org/2017/757}.

\bibitem{Tanner81}
M.~Tanner, ``A recursive approach to low complexity codes,'' \emph{{IEEE}
  Trans. Inf. Theory}, vol.~27, no.~5, pp. 533--547, Sep. 1981.

\bibitem{RL09}
W.~Ryan and S.~Lin, \emph{Channel codes -- {Classical} and modern}.\hskip 1em
  plus 0.5em minus 0.4em\relax New York, NY, USA: Cambridge University Press,
  2009.

\bibitem{Gallager63}
R.~Gallager, \emph{Low-density parity-check codes}.\hskip 1em plus 0.5em minus
  0.4em\relax Cambridge, MA, USA: MIT Press, 1963.

\bibitem{huffman2010fundamentals}
W.~C. Huffman and V.~Pless, \emph{{Fundamentals of Error-Correcting
  Codes}}.\hskip 1em plus 0.5em minus 0.4em\relax Cambridge University Press,
  2010.

\bibitem{FossorierBF2005}
N.~Miladinovic and M.~P. Fossorier, ``{Improved Bit-Flipping Decoding of
  Low-Density Parity-Check Codes},'' \emph{{IEEE} Trans. Inf. Theory}, vol.~51,
  no.~4, pp. 1594--1606, Apr. 2005.

\bibitem{mitzenmacher1998note}
M.~Mitzenmacher, ``{A Note on Low Density Parity Check Codes for Erasures and
  Errors},'' \emph{SRC Technical Note}, vol. 1998, no.~17, 1998.

\bibitem{chung2001design}
S.-Y. Chung, G.~D. Forney, T.~J. Richardson, and R.~Urbanke, ``On the design of
  low-density parity-check codes within $0.0045$ db of the {Shannon} limit,''
  \emph{{IEEE} Commun. Lett.}, vol.~5, no.~2, pp. 58--60, Feb. 2001.

\bibitem{Lentmaier05}
M.~Lentmaier, D.~V. Truhachev, K.~S. Zigangirov, and D.~J. Costello, ``An
  analysis of the block error probability performance of iterative decoding,''
  \emph{{IEEE} Trans. Inf. Theory}, vol.~51, no.~11, pp. 3834--3855, Nov 2005.

\end{thebibliography}

\end{document}